# Multimode semiconductor laser: *quantum versus classical behavior*

M. Lebedev[1,2a)], A. Demenev[1], A. Parakhonsky[1] and O. Misochko[1,2]

[1]*Institute of Solid State Physics, Russian Academy of Sciences, Moscow region, Chernogolovka, 142432, Russia*

[2]*Moscow Institute of Physics and Technology (State University), Moscow region, Dolgoprudny, 141700, Russia*

Temporal correlations of radiation intensities of a multimode Fabry-Perot semiconductor laser are investigated. Strong intensity correlations with a fixed phase shift between different longitudinal modes of the laser are revealed. The second $g^{(2)}$ and third $g^{(3)}$ order intensity correlation functions are studied to clear the character of the intermodal coupling.

## I. INTRODUCTION

Multimode semiconductor lasers show a variety of highly nontrivial effects, including chaotic pulse generation, two-photon interference, nonlinear gain, optical feedback and many others [1]. Some of them can be well understood within classical nonlinear optics, whereas the others are true quantum optics effects, meaning that their explanation cannot be given without quantum theory of light. To distinguish between quantum and classical behavior is not easy, because almost every quantum effect has its own classical counterpart as quantum optical coherent states have many properties in common with classical coherent light fields. A typical example is photon superbunching that is very similar to chaotic pulse generation. We observed recently strong intensity correlations for longitudinal modes of a continuous wave multimode semiconductor laser [2,3] which can be well described in either the frame of totally quantum consideration involving back action of the measurement process on the emitting system or, alternatively, as a result of a chaotic generation of classical pulses. Trying to distinguish between these two options we measured a third order intensity

---

[c)] E-mail: lebedev@issp.ac.ru

correlation function $g^{(3)}(\tau)$ and the results of these measurements are presented in this paper.

It is well known that temporal quantum correlations, in analogy to the spatial correlations, quantified by the Bell inequalities [4], allow separating classical from quantum behavior. Indeed, when the Leggett–Garg inequality [5], involving correlations of measurements on a system at different times, is violated the time evolution cannot be understood classically. Generally speaking, the $g^{(3)}(\tau)$ correlator measurement exploits in principle the same idea as the Leggett-Garg inequalities. If we observe a quantum mechanical system, the observation disturbs its free evolution and causes the dependence of the probability of results of a subsequent measurement on the fact of a previous observation.

## II. EXPERIMENTAL RESULTS AND DISCUSSION

The spectrum of our continuous multimode semiconductor Fabry-Perot laser shown in Fig.1a consists of the large number (approximately 35) longitudinal modes grouped into broad peaks corresponding to different transverse modes. Every transverse mode contains up to 7 longitudinal ones with the longitudinal mode spacing of 0.11 nm. Using a double monochromator, one could select a desired spectral range to study the intensity correlation function either for a single longitudinal mode, or for the chosen number of such modes. This selection could be easily done because the monochromator could operate either in the dispersion subtraction mode, which gives the possibility to study the dependence of intensity correlations on the number of longitudinal modes transmitted through the monochromator, or in the dispersion addition mode in which a single longitudinal mode can be studied. Moreover, some modernization of our setup gave us the possibility to study intensity correlations between different longitudinal modes. In this case the exit slit of the monochromator was removed and two fibers were positioned in the plane of the exit slit. Each fiber could be moved independently and captured the radiation of only one desired longitudinal mode (dispersion of the monochromator was high enough to fill the entrance aperture of each fiber with radiation of one longitudinal mode only).

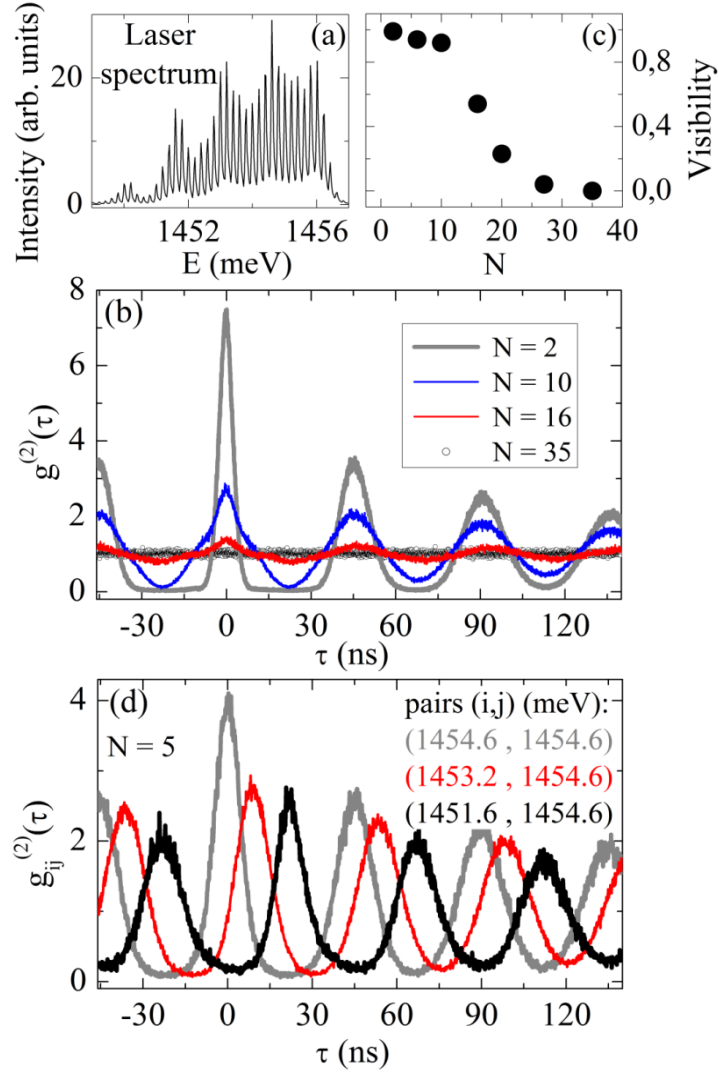

FIG.1. a) The spectrum of our semiconductor FP-laser, b) $g^{(2)}(\tau)$ intensity correlator for varying number N of simultaneously recorded longitudinal modes, c) Visibility of the $g^{(2)}(\tau)$ correlator as a function of N, d) $g_{i,j}^{(2)}(\tau)$ is the cross-correlation function of the intensities of given pairs of longitudinal laser modes.

We started with measurements of the $g^{(2)}(\tau)$ intensity correlation function by an ordinary Brown-Twiss setup and found that the result depends critically on the number of simultaneously selected modes [2, 3]. If all the laser modes were transmitted through the monochromator, or the Brown-Twiss interferometer was illuminated with the laser directly bypassing the monochromator, no intensity correlations were observed, see Fig.1b. In other words, if one uses a monochromator with a rather low spectral resolution one will get a "monochromatic" laser generation line that exhibits Poissonian photon statistics. This result looks very natural, because the monochromatic laser light is known to be in a coherent state which gives no correlations between photons. Theoretical considerations of a multimode laser generation contain frequently an assumption that every mode of such laser is in its own coherent state, which is arbitrarily shifted in phase against the coherent states of the other modes. Our measurements [2, 3] showed that it is not our case. The $g^{(2)}(\tau)$ intensity correlation function for a single longitudinal mode shows oscillations between superbunching and antibunching regimes with a surprisingly high visibility $V = 0.98$:

$$V = \frac{g^{(2)}_{max} - g^{(2)}_{min}}{g^{(2)}_{max} + g^{(2)}_{min}} \quad (1)$$

and very long correlation times (see Fig.2). Increasing the number of simultaneously detected longitudinal modes leads to a monotonic decrease of the visibility (see Fig.1b), so in the limit of full spectrum detection the visibility tends to zero as shown in Fig.1c, reproducing again the Poissonian statistics. It is straightforward to conclude that this result can take place only if the intensity correlation between different longitudinal modes is strong and has a fixed phase shift which depends on the chosen pair of longitudinal modes. The cross-correlation intensity fluctuation measurements prove that this is really the case as shown in Fig.1d. Note, that it has been demonstrated experimentally that for different free-running multi-longitudinal-mode semiconductor lasers, the modal outputs display large amplitude antiphase oscillations that leave constant the total intensity emitted by the laser [6].

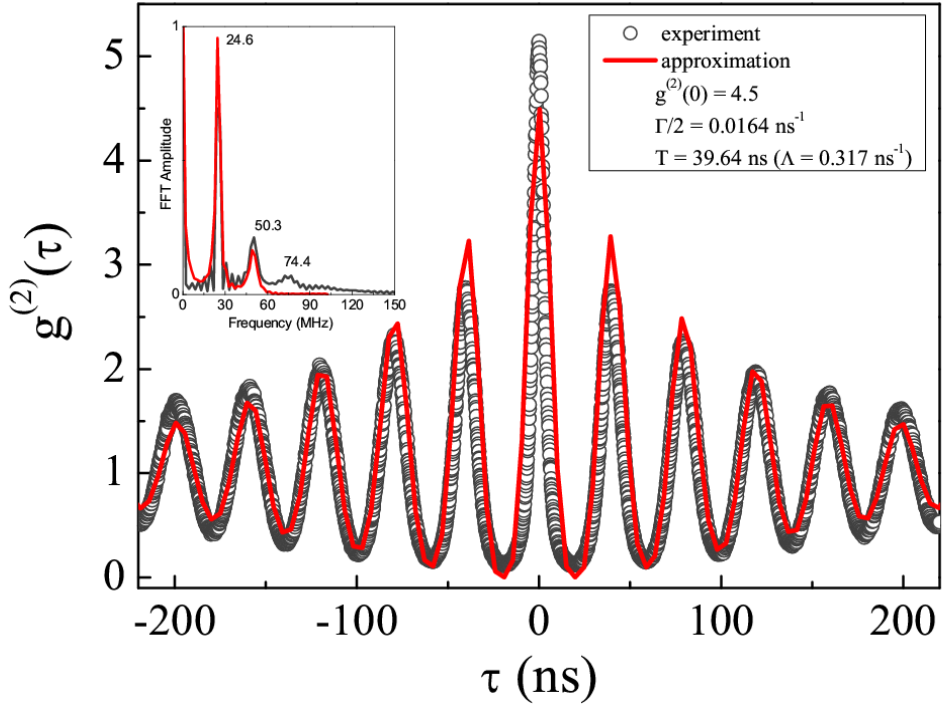

FIG.2. $g^{(2)}(\tau)$ correlator of a single longitudinal laser mode (open cycles) and its approximation with the Eleuch's formula from [7] (the red curve). The insert shows corresponding Fourier transforms in the same color.

All these results look very interesting but say nothing about the nature of the correlations. The only thing which we could definitely say is that these correlations are not caused by the pump current fluctuations of the laser diode, because a direct measurement of the pump current stability in a nanosecond time scale showed the absence of any detectable fluctuations of the pump. The most natural explanation of our results could be strong optical interaction between the longitudinal modes of the laser resonator due to a great variety of possible nonlinear interaction mechanisms inherent to semiconductor lasers [6]. In this case, the correlations should be classical, that means that our measurement process (the single photon detection) does not affect the next photon emission.

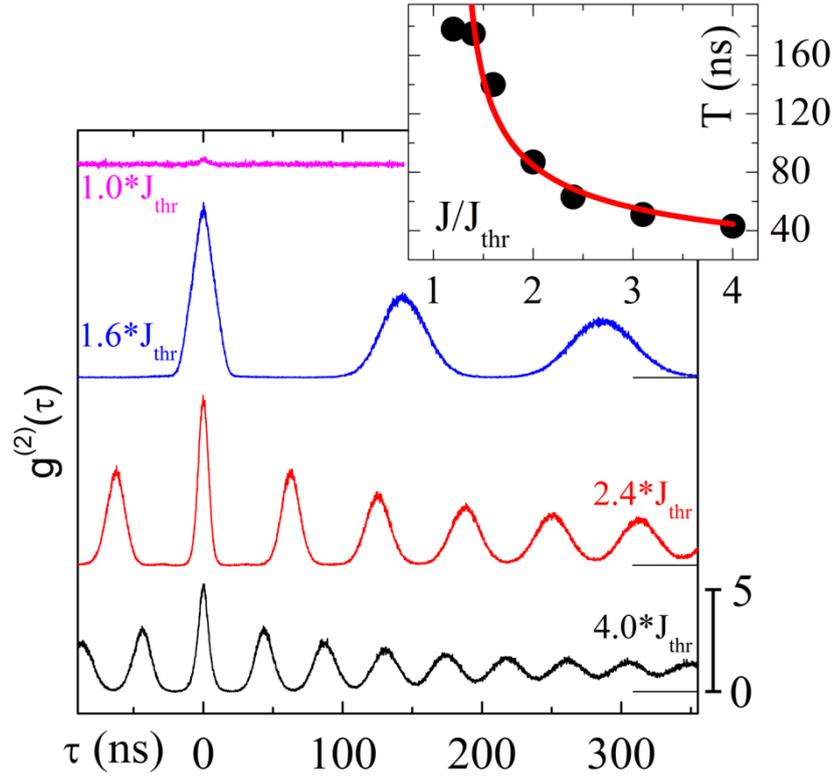

FIG.3. The intensity correlator $g^{(2)}(\tau)$ of a single longitudinal laser mode at different pumping currents ($J_{thr}$ denotes the laser threshold current). The insert: dependence of the period of oscillations on the pumping current normalized to the threshold current. The red line is the approximation of experimental data with the inverse square root curve.

On the other hand, a totally quantum mechanical explanation is also possible [7]. A semiconductor laser can be represented as an ensemble of two-level systems strongly interacting with a light mode. Two-photon interference effects recently observed in a cw multimode semiconductor laser show that such lasers can generate nonclassical light [8]. A strong coherent interaction between the light and the electron-hole system in a semiconductor laser is a fundamental principle of the semiconductor laser theory. This interaction leads to the Rabi oscillations between the light mode and the electron-hole system. The observed oscillating character of $g^{(2)}(\tau)$ correlator could be explained by Rabi oscillations because the period of these oscillations has an inverse square root dependence on the pumping current as shown in Fig.3. Quantum mechanical consideration of the phenomenon [7] leads to the expression for the $g^{(2)}(\tau)$ function that fits well the experimental data as shown in Fig.2. The reason for the lack of agreement between the experimental data and the fitting is easily seen by comparing the Fourier transforms of the experimental correlator and the fitting result shown in the inset in Fig.2. In the first, the Fourier transform in addition to the fundamental and second harmonic of the modulation frequency, contains also a third harmonic, while the third harmonic is absent in the Fourier transform of the fitting result. The latter is a consequence of approximation of the system state vector in [7] with a two-quantum state neglecting higher harmonics. We decided to measure the $g^{(3)}(\tau)$ intensity correlation function hoping that these measurements could provide more information about the nature of correlations observed. As has been mentioned the $g^{(3)}(\tau)$ correlator measurement exploits similar ideas as the Leggett-Garg inequalities [5]. Namely, if we observe a quantum mechanical system, the observation disturbs its free evolution and causes the dependence of the probability of results of a subsequent measurement on the fact of previous observation.

Our experimental setup to measure the $g^{(3)}(\tau)$ correlator is shown in Fig.4. The radiation of a single longitudinal mode was fed into a fiber and split with the aid of two beamsplitters between three APD (avalanche photodiodes) detectors. Optical paths from the monochromator to the photodetectors were approximately equal (within 1 meter). The signal of the first APD (D1) could trigger the gated start input of the time-to amplitude converter (TAC), to which the signal from D2 was connected. The gating pulse of D1 was 8ns long and fed to the gating input of the TAC through a variable delay line. If TAC received a start pulse during the 8ns gate it began to operate waiting for the stop pulse from the detector D3. As a result, our setup was measuring the $g^{(3)}(\tau)$ correlator:

$$g^{(3)} = \frac{\langle I(t_1)I(t_2)I(t_3)\rangle}{\langle I\rangle^3} \qquad (2)$$

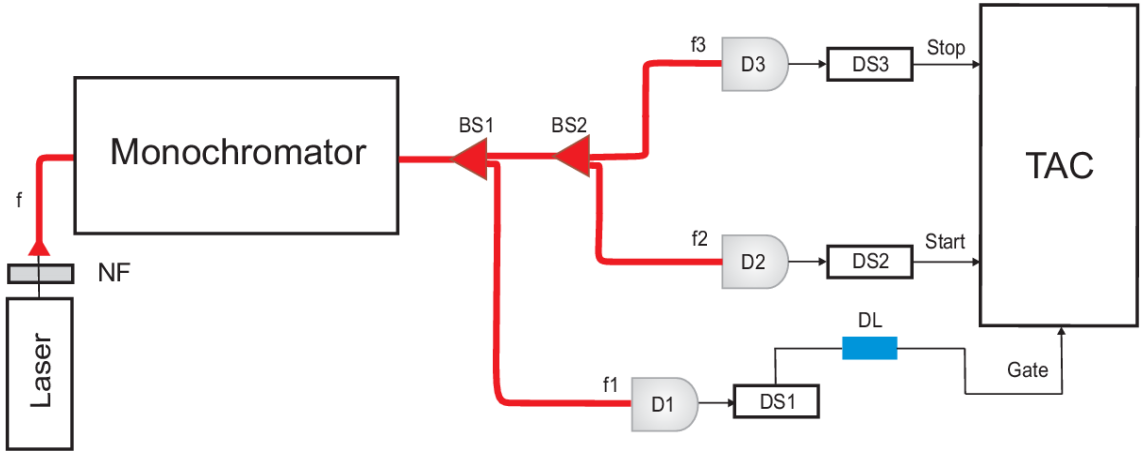

FIG. 4. Schematic of the experimental setup. BS1 and BS2 are fiber-Y-splitters, f, f1, f2 and f3 are optical fibers, NF is a neutral density filter, D1, D2 and D3 are single photon detectors (silicon avalanche photodiodes (APD)), DS1, DS2 and DS3 are discriminators, DL is a delay line, TAC is a time-to amplitude converter.

Due to the stationarity of the measured process this correlator depends on time differences only

$$g^{(3)} = g^{(3)}(t_2 - t_1, t_3 - t_2) = g^{(3)}(\Delta, \tau) \qquad (3)$$

Typical experimental results are shown in Figs.5,6 and 7. The $g^{(3)}(\tau)$ correlator demonstrates oscillations with high visibility similar to the $g^{(2)}(\tau)$ correlator and has two main maxima: one at $\tau = -\Delta$ which corresponds to $t_1 = t_3$ and the second at $\tau = 0$, that is, at $t_2 = t_3$. Such behavior looks quite natural, because the intensities of light which illuminates the detectors are equal for the detectors D1 and D3 in the first case, and for the detectors D2 and D3 in the second one. If $\Delta$ is large enough (significantly exceeds the correlation time of $g^{(2)}$) one can expect that the correlations between photons 1 and 2 totally vanish and the $g^{(3)}(\tau)$ correlator should reduce to the product of the $g^{(2)}(\tau)$ correlators. This follows from the fact that the correlations between photons 1 and 3 are registered in a time interval which is well separated in time from the time where correlations between photons 2 and 3 are registered and one has essentially two independent $g^{(2)}(\tau)$ measurements.

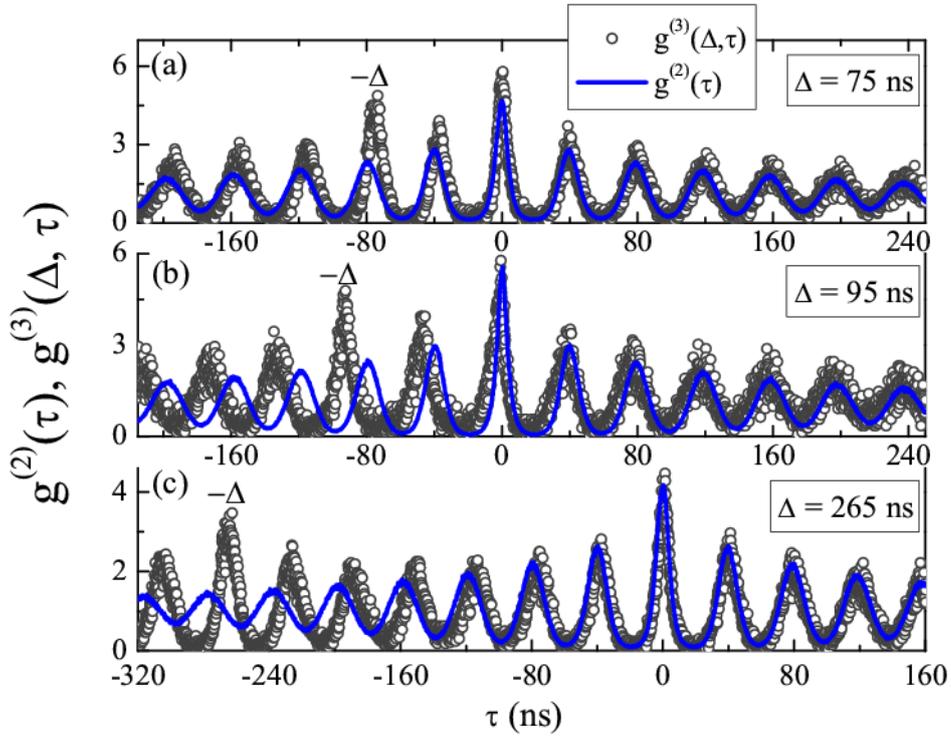

FIG. 5. Triple correlator $g^{(3)}(\tau)$ as a function of delay $\tau$ (open cycles) at different time shifts of the strobe pulse $\Delta$ and the $g^{(2)}(\tau)$ correlator (blue lines).

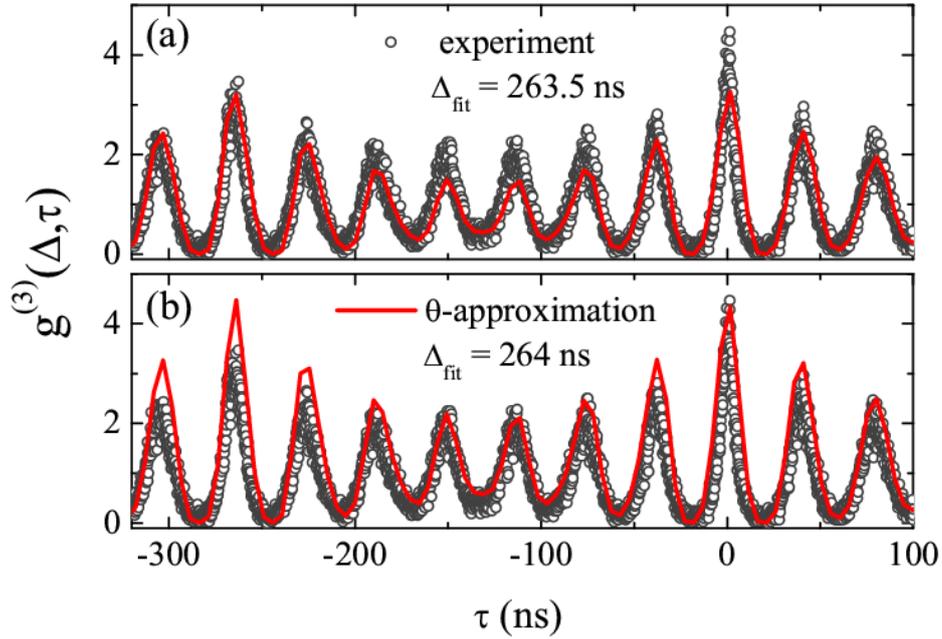

FIG.6. Experimental values of the $g^{(3)}(\tau)$ correlator (open cycles) and theoretical approximations (red lines) for large $\Delta$: a) "Classical": Eq. (5) and b) "quantum" : Eq. (8). The only parameter used for the fitting was the $\Delta$ value. The experimentally determined value was 265 ns.

The product of two correlators preserves the high visibility of individual $g^{(2)}(\tau)$ correlators. According to this consideration in the limit of large $\Delta$, the non-normalized triple correlator can be defined as:

$$G^{(3)}(\Delta, \tau) = g^{(2)}(\Delta)g^{(2)}(\tau)g^{(2)}(\tau + \Delta) \tag{4}$$

At large $\Delta$ we have $g^{(2)}(\Delta) \to 1$. Normalizing Eq. (4) with $g^{(2)}(\Delta)$ one gets for $g^{(3)}(\Delta, \tau)$:

$$g^{(3)}(\Delta, \tau) = g^{(2)}(\tau)g^{(2)}(\tau + \Delta) \tag{5}$$

This expression should hold for any classical chaotic pulses at any $\Delta$ because the measurement process does not affect the photon registration probability. But Eq. (5) obviously fails in the quantum mechanical case where every measurement prepares a new state of the system and all correlations with a previous state are erased. This can be illustrated by the work of Eleuch [7]. He considered an ensemble of two-level systems (excitons in a microcavity) which is resonantly excited with a weak laser radiation incident normally to the microcavity plane. The low power of the pumping laser gives the possibility to consider the exciton-light system inside the cavity as being approximately in a pure quantum mechanical state, which evolution is governed by an effective non-hermitian Hamiltonian [7]. Note, that in a case of a multimode semiconductor laser, we also have an ensemble of two-level systems (electron-hole pairs) interacting with laser light inside the laser resonator and pumped with an injection current. According to [7], the initial state of the exciton-photon system is supposed to be a steady state described with the state vector $\Psi_{ss}$ which can be expanded in a superposition of tensor product of pure excitonic and photonic states. Detection of the first photon causes the system to undergo a quantum jump: $a\Psi_{ss} \to \varphi$ where $a$ is the photon annihilation operator. The time evolution of the state vector $\varphi$ is governed by the same Hamiltonian as the time evolution of $\Psi$ which gives the possibility to calculate $g^{(2)}(\tau)$:

$$g^{(2)}(\tau_1) = \frac{\langle \varphi(\tau_1)|a^+ a|\varphi(\tau_1)\rangle}{\langle a^+ a \rangle} \tag{6}$$

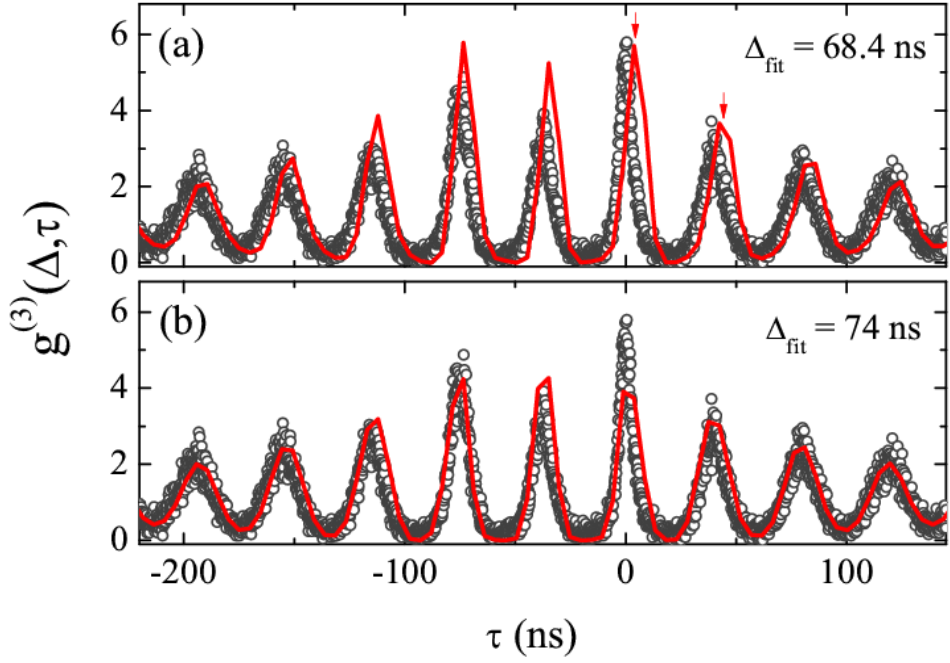

FIG.7. Symbols and curves same as in FIG. 6 but for Δ=75 ns. "Quantum" approximation (b) fits the experimental data evidently better.

Returning to the triple correlator, we see that in the case of large enough Δ the system will reach the steady state between the moments of registration of photons 1 and 2, so we get again the result Eq. (5) as for the classical chaotic optical pulses. But for small enough Δ, one has to exploit the same procedure as for calculating $g^{(2)}(\tau_1)$:

$$\tilde{g}^{(2)}(\tau_2) = \frac{\langle \chi(\tau_2)|a^+a|\chi(\tau_2)\rangle}{\langle a^+a \rangle}, \quad (7)$$

where $a\varphi \to \chi$ ($\tau_2 > \tau_1$). We see that in principle $\tilde{g}^{(2)}(\tau_2) \neq g^{(2)}(\tau_1)$ but they may be very close if the steady state of the system is more than a two quantum state, because their spectra will differ only with an amplitude of a high order harmonic. A more evident consequence of quantum mechanical picture is the erasing of previous correlations after a quantum jump of the system due to reduction of its state vector into a new quantum state. This means that in the quantum mechanical case one has to replace Eq. (5) with:

$$g^{(3)}(\Delta,\tau) = \theta(\tau)g^{(2)}(\tau) + \frac{\theta(-\tau)\theta(\Delta+\tau)g^{(2)}(\tau)g^{(2)}(\Delta+\tau)}{g^{(2)}(\Delta)} + \theta(-\Delta-\tau)g^{(2)}(\Delta+\tau) \quad (8)$$

We compared our results with Eqs. (5) and (8) (see Figs.6 and 7) and found that for large $\Delta$ (see Fig.6) both expressions demonstrate reasonable agreement with experiment. However, for small $\Delta$ (see Fig.7) Eq.(8) fits the experimental data much better than Eq. (5). In the fitting procedure we used for the $g^{(2)}(\tau)$ correlator the analytical expression of Eleuch [7] with parameters fixed in the fitting shown in Fig.2. The only variable parameter used in the $g^{(3)}(\tau)$ fitting procedure was $\Delta$.

### III. CONCLUSION

Summarizing, we can say that quantum two-photon interference effects [8] and highly nontrivial $g^{(2)}(\tau)$ correlator behavior [2,3] of used multimode semiconductor laser light may be the signature of a macroscopic light-matter quantum state, because the $g^{(3)}(\tau)$ correlator measurements reveal some influence of the photon detection on the probability to detect the next photon. The best way to test the quantum nature of the light–matter system would be of course a direct verification of the Leggett - Garg inequalities but it requires the introduction of some kind of dichotomous variables which should be done for our case. That is the Leggett–Garg inequalities should be constructed from time correlation functions of a series of measurements of some dichotomous variable carried on a considered system.

This work was in part supported by the Russian Foundation for Basic Research (No. 17-02-00002) and by the Government of the Russian Federation (Agreement No. 05.Y09.21.0018).